\documentclass[12pt,a4paper]{article}

\usepackage{amsmath}
\usepackage{amssymb}

\usepackage[latin1]{inputenc}

\usepackage{graphicx}
\usepackage{float}
\usepackage{subfigure}
\usepackage[caption=false]{subfig}  

\usepackage{cite}
\usepackage{url}
\usepackage{algorithm}  
\floatname{algorithm}{Alg.}  
\usepackage{algorithmic} 
\usepackage{pgf}
\usepackage{pgfplots}
\usepackage{tikz}
\usepackage{tikz-inet}
\usetikzlibrary{patterns}
\usetikzlibrary{arrows,automata,plotmarks}
\usepgflibrary{shapes} 
\usepgflibrary[shapes] 
\usetikzlibrary{shapes} 
\usetikzlibrary[shapes] 

\usepackage{calc}
\usepackage[latin1]{inputenc} 
\usepackage{tabularx}
\usepackage{fullpage}
\usepackage{ifpdf}
\usepackage{xcolor}
\usepackage{eurosym}

\makeatletter
\renewcommand{\fnum@figure}{\small\textbf{\figurename~\thefigure}}
\renewcommand{\fnum@table}{\small\textbf{\tablename~\thetable}}
\makeatother

\usepackage{color}

\newcommand{\define}{:=}
\newcommand{\ber}{B}
\newcommand{\hop}{h}
\newcommand{\frr}{\alpha}
\newcommand{\nfrag}{m}
\newcommand{\linkdata}{{\mathbf D}}
\newcommand{\linkack}{{\mathbf A}}
\newcommand{\linkrmax}{r}
\newcommand{\per}{F}
\newcommand{\HS}{{\mathbf H}}
\newcommand{\PEE}{Q}
\newcommand{\EE}{{\mathbf E}}
\newcommand{\Seg}{{\mathbf S}}
\newcommand{\TEE}{{\mathbf I}}
\newcommand{\pfail}{P_{\text{fail}}}
\newcommand{\ppartial}{P_{\text{partial}}}
\newcommand{\psucc}{P_{\text{succ}}}
\newcommand{\N}{\linkdata}
\newcommand{\K}{{\mathbf K}}

\newcommand{\comb}[2]{\left(\!\!\!\begin{array}{c}#2\\#1\end{array}\!\!\!\right)}

\begin{document}

%
%
%


\title{
TCP over low-power and lossy networks: tuning the segment size to minimize energy consumption
\\ ~ \\
{\Large {\it TELECOM Bretagne Research Report}}
\\
}


\author{\\
\\
Ahmed Ayadi,  Patrick Maillé and David Ros \\
{\normalsize TELECOM Bretagne} \\
{\normalsize Networks, Security and Multimedia (RSM) department} \\
{\normalsize Rue de la Châtaigneraie, CS17607} \\
{\normalsize 35576 Cesson Sévigné Cedex, France} \\
{\normalsize e-mail: \texttt{\{ahmed.ayadi,patrick.maille,david.ros\}@telecom-bretagne.eu}}
}

\date{}

\maketitle

\newpage

\section*{Abstract}
Low-power and Lossy Networks (LLNs), like wireless networks based upon the IEEE 802.15.4 standard, have strong energy constraints, and are moreover subject to frequent transmission errors, not only due to congestion but also to collisions and to radio channel conditions.

This paper introduces an analytical model to compute the total energy consumption in an LLN due to the TCP protocol. The model allows us to highlight some tradeoffs as regards the choice of the TCP maximum segment size, of the Forward Error Correction (FEC) redundancy ratio, and of the number of link-layer retransmissions, in order to minimize the total energy consumption.

\section*{Keywords}

Multi-Hop Wireless Networks, Low-Power and Lossy Networks, Energy Efficiency, TCP.




\newpage

\tableofcontents

\newpage
\listoffigures
\listoftables

\newpage
\section{Introduction\label{sec:introduction}}

IP-based Low-power and Lossy Networks (LLNs) are composed of (possibly many) nodes which are subject to strong constraints on power and computing capabilities. Besides, link-layer technologies like IEEE 802.15.4 \cite{ieee802154}, often used for building wireless LLNs, impose very low limits on link frame sizes.

Many usual applications for LLNs (e.g.,\ wireless sensor networks) do not always require full reliability. However, there exist several use-cases for LLNs where the reliability of data delivery can be critical, like over-the-air software updates \cite{1413478,3478765,1340217}. In such cases, the Transmission Control Protocol (TCP) \cite{rfc:tcp} is a common choice for ensuring an end-to-end reliable transport in IP-based LLNs.

TCP is the most used transport protocol in IP networks. It provides a reliable transfer service for all kinds of applications. TCP can manage data that is damaged, duplicated, lost, or delivered out of order. This is achieved by assigning a sequence number to each octet transmitted and requiring a positive acknowledgement (ACK) from the recipient. If the TCP ACK segment is not received within the timeout interval, the data segment is retransmitted.

This paper focuses on the energy cost of reliability when TCP is used in multi-hop wireless LLNs. In what follows, we will illustrate some main issues by considering the case of TCP in IPv6-enabled LLNs based on the 6LoWPAN protocols \cite{rfc:lowpan}.

Many parameters affect the energy consumption of a reliable transport protocol, like channel conditions (e.g., wireless losses, collisions), the level of link-layer reliability, the number of links/hops, or the maximum size of link-layer frames (e.g., 127 bytes for IEEE 802.15.4). The length of such frames may require the fragmentation of IPv6 datagrams before sending them. The IETF 6LoWPAN working group has defined a new protocol layer between the IPv6 and the medium access control (MAC) layers. The 6LoWPAN layer compresses the IPv6 header and fragments IPv6 packets into short MAC frames. So, if the size of a TCP segment exceeds the maximum allowed length, the 6LoWPAN layer fragments it to make it fit into link-layer frames. Therefore, sending short TCP segments does not require any fragmentation. However, the use of a small maximum segment size\footnote{The TCP MSS option refers to the maximum amount of data a TCP segment can carry, i.e., excluding the TCP header and any options that might be present.
} (MSS) increases the number of TCP segments and, thus, the number of TCP acknowledgements and the corresponding MAC frames sent. Besides, with small MSS values the overhead due to TCP headers becomes larger; this, coupled with the small size of MAC frames, may result in a very low protocol efficiency.


This paper introduces a mathematical model aimed at predicting the energy consumed by the wireless nodes of an LLN in a bulk-data transfer scenario, with TCP used as a reliable transport layer. The model estimates TCP energy performance based on the bit error rate, the maximum number of retransmissions at the  link layer, the number of hops between the sender and the receiver, the amount of FEC, and the TCP maximum segment size.

The proposed model allows us to study the tradeoffs involved in sending short versus long TCP segments. We assume that the energy consumed in a data transfer depends mainly on the number of bits sent. Thus, our analytical model estimates the number of bits sent by all nodes in the network, taking into account the cumulated cost of all link-layer transmissions. Indeed, the number of MAC frames sent can be large with respect to the initial amount of data to transmit, because of link-layer retransmissions, the redundancy added for Forward Error Correction (FEC), and also due to end-to-end (i.e., transport layer) retransmissions when a TCP segment is lost before reaching the receiving node.

We apply the model to study the energy efficiency of TCP over an LLN using 6LoWPAN and 802.15.4 protocols, and study the effect of the TCP segment size, of the FEC  redundancy ratio,  and of the maximum link layer retransmission on the total energy consumption.

The remainder of the paper is organized as follows. Section~\ref{sec:related_work} gives a brief overview of the related work in the area of TCP performance in LLNs, and in multihop wireless networks in general. Section~\ref{sec:energy_models} presents the derivation of our analytical model. In Section~\ref{sec:analytical_simulation_result}, we
apply the analytical results to derive the best TCP segment size strategy in terms of energy consumption. Finally, in Section~\ref{sec:conclusion}, we give our conclusions and provide directions for future work.

\section{Related work}
\label{sec:related_work}

There exist many studies of the performance of TCP in multi-hop wireless networks (see for example~\cite{521478} and references therein), though many of them are concerned mostly with parameters like TCP throughput, and with protocol enhancements aimed at improving such performance metrics. Another amount of work focuses on the computational cost of TCP (e.g., \cite{347765}), or on the comparison of the energy efficiency for different TCP congestion control algorithms (like SACK, Tahoe, Reno and NewReno) over wireless networks (e.g., \cite{4149413,1143650,383765}).
In ~\cite{FuLZLZG05}, Fu et al. study in the impact of the MAC layer protocol on the TCP throughput over wireless multi-hop networks. The authors show that, in contrast to wired networks,  the TCP throughput does not increase by increasing the window size, and that (under the conditions of the study) better performance is achieved with a window size equal to the number of hops divided by four.

In~\cite{34854}, Lilakiatsakum and Senevirane propose
an energy-efficiency metric to compare the performance of different versions of TCP.
The metric is the ratio between the amount of bits sent by all nodes and the size of the application data. In this work, we adopt a variant of that metric (the total amount of bits sent); we use a fixed value for the application data size in all our numerical and simulation scenarios.

Bansal et al.~\cite{1143381} propose an analytical model to compute the energy consumed for carrying a TCP flow over a multi-hop wireless network. The authors assume that all wireless links have the same packet error rate $P$ and the same transmission power. They compute the energy spent by all nodes for sending a single TCP segment. The energy consumption  is then a function of the packet error rate, the number of hops, and the maximum number of link-layer retransmissions. However, \cite{1143381} does not take into account the cost of sending link-layer acknowledgements, nor the cost of transport-layer acknowledgements, which we add in the model introduced here. Besides, we explicitly consider the impact of lower-layer fragmentation and of error correction, as well as different per-link error rates on the energy-efficiency of TCP.

Barman et al.~\cite{597603204} and Gallucio et al.~\cite{12179506} present an analytical study of a TCP optimization problem in an hybrid wired/wireless network where the last hop is a wireless link. The authors of both papers define an utility function which is the ratio of the throughput to the cost of a TCP connection. Our work completes these studies by introducing a multi-hop model for computing the energy consumption. Note that in this paper we are not interested in TCP throughput, because data rates are a secondary concern in LLNs; instead, we focus mainly on the energy costs of a TCP connection over multiple lossy links.

\section{TCP energy consumption model}
\label{sec:energy_models}

This section introduces a mathematical model to estimate the energy consumed by a TCP transmission in a wireless LLN. We assume that such energy mainly corresponds to data emission and reception, and thus directly depends on the number of bits sent by all nodes. 

We therefore compute the expected total amount of bits sent for a successful end-to-end TCP data transmission, as a function of several network parameters, namely: the bit error rate, the link-layer maximum number of attempts, the FEC  redundancy ratio, the number of hops between the source and receiver TCP hosts, and the TCP maximum segment size.

For the convenience of the reader, Table~\ref{tab:variable_energy} lists most of the variables used in this paper.

\begin{table}[t]
\caption{Notations used in this paper; capital italics letters correspond to probabilities, bold letters to (expected) numbers of bits.\label{tab:variable_energy}}
\centering
\begin{tabular}{|c|p{12cm}|}
\hline \textbf{Variable} &  \textbf{Definition}\\ 
\hline $\hop{}$ & Number of hops between source and destination\\ 
\hline $\linkrmax{}$ & Maximum number of transmission attempts at the link layer\\
\hline $\nfrag$ & Number of fragments corresponding to a single TCP segment (due to link layer fragmentation)\\
\hline $\frr$ & FEC redundancy ratio\\
\hline $\ber{}$ &  Bit error rate\\ 
\hline $\pfail$ & Probability of a failure in a link-layer transmission attempt of a data frame\\
\hline $\ppartial$ & Probability of a link-layer data frame being correctly received and the corresponding acknowledgement being lost\\
\hline $\psucc$ & Probability of a successful link-layer transmission attempt (data+acknowledgement are correctly received)\\
\hline $\per{}$ & Probability that a destination node does not receive a link layer data frame after $r$ attempts\\
\hline $\PEE{}_{s}$ & Probability of an end-to-end packet transmission success\\
\hline $\PEE{}_{f}$ & Probability of an end-to-end packet transmission failure\\
\hline $\linkdata{}$ & Link-layer data frame size\\
\hline $\linkack{}$ & Link-layer acknowledgement frame size\\
\hline $\HS{}_{f}$ & Expected number of  bits sent after $r$ attempts knowing that the (one-hop) transmission has failed\\
\hline $\HS{}_{s}$ & Expected number of bits sent within $r$ attempts knowing that the (one-hop) transmission has succeeded \\
\hline $\EE{}_{f}$ & Expected number of bits sent for an end-to-end packet transmission knowing that it has failed\\
\hline $\EE{}_{s}$ & Expected number of bits sent for a successful end-to-end packet transmission knowing that it has succeeded\\
\hline $\Seg$ & Average number of bits sent for successfully transmitting a TCP segment\\
\hline 
\end{tabular} 
\end{table}

\subsection{Link layer: one-hop model}
\label{sec:link_layer_models}

We first focus on modeling the link-layer energy consumption, considering a CSMA-CA network with error correction control techniques combining Automatic Repeat Request (ARQ)~\cite{rfc:arq} and Forward Error Correction (FEC).

\subsubsection{Link layer mechanisms}

Standard ARQ uses the cyclic redundancy check (CRC) error-detecting code that is added to the data: The receiver uses the error-detecting code number to check the integrity of the received data. After receiving a correct frame, the receiver replies by an ACK. If the sender does not receive any ACK before a timeout\footnote{Remark that we do not consider time issues in this model, thus, only losses can lead to retransmissions.}---because either the original message or the ACK is lost, or they contain errors---, the sender retransmits again the same message. If the receiver sees that the frame is damaged the receiver discards it and does not send an ACK. The ARQ algorithm continues until the sender receives an ACK or exceeds a predefined number of attempts~$\linkrmax{}$.

ARQ is the algorithm most frequently used by link-layer protocols 
to reduce the packet error rate. However, if the wireless network becomes very lossy, ARQ would increase the transmission delay between the source and the receiver.
Abrupt increases of the end-to-end delay increase the round-trip time and may lead to a spurious TCP timeout. This can deteriorate the TCP performance.

An orthogonal approach consists in applying Forward Error Correction (FEC). FEC is a good solution for decreasing the packet error rate. 
The main idea of FEC is to add redundancy to the original frame, to allow the destination node to detect and correct some bit errors.
In our case, if the size of a network datagram is greater than the maximum transmission unit (MTU) of the link layer, the datagram is divided into fragments of length $\K$ bits, and the FEC algorithm adds $(\frr \times \K)$ redundancy bits to form a frame of length $\N$. 
The ratio $\frr=\frac{\N-\K}{\K}$ between the amount of redundancy added by FEC and the original frame length is called the \emph{redundancy ratio}.

In what follows we will consider an error-correction method like the well-known Reed-Solomon~\cite{reed60polynomial} (RS) algorithm. 
By adding $\N-\K$ bits to the $\K$ data, the RS algorithm can correct up to $(\N-\K)/2$ bits.

\subsubsection{Performance of one-hop transmissions}

\begin{figure}[ht]
\begin{center}
\subfigure[Failure.]{
	\begin{tikzpicture}
	\draw [very thick] (-1,-1) -- (-1,1)  node [above] {Sender} ;
	\draw [very thick] (1,-1) -- (1,1) node [above] {Receiver};
	\draw [thick] [->] (-1,0.5) -- node[midway,sloped,above] {Data frame} (0.75,0.3125) ;
	\node [cross out,draw=red] at (0.75,0.3125) {.};
	\end{tikzpicture}
} 
\subfigure[Partial failure.]{
	\begin{tikzpicture}
	\draw [very thick] (-1,-1) -- (-1,1) node [above] {Sender} ;
	\draw [very thick] (1,-1) -- (1,1) node [above] {Receiver};
	\draw [thick] [->] (-1,0.5) -- node[midway,sloped,above] {Data frame} (1,0);
	\draw [thick] [->] (1,-0.2) -- node[midway,sloped,above] {Ack frame} (-0.8,-0.33);
	\node [cross out,draw=red] at (-0.8,-0.33) {.};
	\end{tikzpicture}
} 
\subfigure[Success.]{
	\begin{tikzpicture}
	\draw [very thick] (-1,-1) -- (-1,1) node [above] {Sender} ;
	\draw [very thick] (1,-1) -- (1,1)  node [above] {Receiver};
	\draw [thick] [->] (-1,0.5) -- node[midway,sloped,above] {Data frame} (1,0);
	\draw [thick] [->] (1,-0.2) -- node[midway,sloped,above] {Ack frame} (-1,-0.5);
	\end{tikzpicture}
} 
\end{center}
\caption{Failure and success scenarios for one link-layer transmission attempt.}
\label{fig:link_layer_losses_success}
\end{figure}
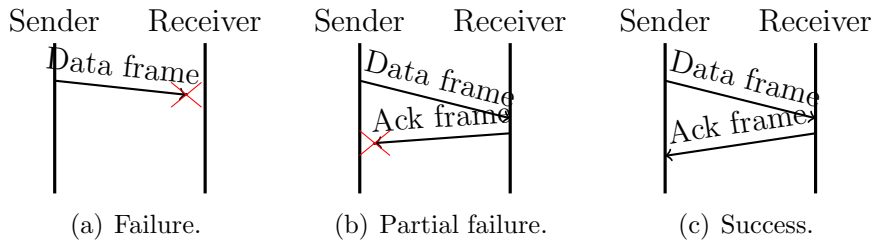

First, we consider here only a transmission between two immediate neighbors, without intermediate nodes. In CSMA-CA, two types of messages are used during data transmission: the data message (i.e., the message that contains the useful data), and the acknowledgement message sent by the receiving node.
Figure~\ref{fig:link_layer_losses_success} shows the three possible cases:
\begin{enumerate}
\item[a.] \emph{Failure:} A failure due to the loss of the data frame. The sender will retransmit the frame.
\item[b.] \emph{Partial failure:} The data frame is correctly received by the receiver, while the acknowledgement frame is lost\footnote{In a multi-hop setting, the receiver would nonetheless relay the frame to the next hop.}. Therefore, the sender will (needlessly) retransmit the data frame.
\item[c.] \emph{Success:} A successful transmission of both the data and the acknowledgement frames.
\end{enumerate} 

The probability of a successful transmission of a link-layer frame depends on the bit error rate, the size of the message, and the redundancy ratio. Here we assume that all bits undergo independent random errors with a given bit error rate $\ber$, each bit being correctly received  with a probability $1-\ber$. 
Moreover, we assume that a frame transmission fails as soon as the number of bits erroneously received exceeds the number of correctable bits.
The probability $\pfail$ of having a \emph{failure} (case (a) in Fig.~\ref{fig:link_layer_losses_success}) is then the error probability for a data frame, i.e.,
\[
\pfail \define 1-\sum_{i=0}^{c}\comb{i}{\linkdata} \ber^i (1-\ber)^{\linkdata-i} 
\]
where $\linkdata$ is the number of bits of a data frame, $c=\frac{\linkdata-\K}{2}=\frac{\frr \K}{2}$ is the number of correctable bits, and for two integers $i$ and $j$, $\comb{i}{j}$ is the number of possibilities of choosing $i$ elements among $j$, i.e., $\comb{i}{j}=\dfrac{j!}{i!(j-i)!}$.

Likewise, denoting by $\linkack$ the size of an acknowledgement frame,  
the probability of a \emph{partial failure}  is the probability that the data frame is correctly received but the acknowledgement frame contains errors:

\[
\ppartial\define (1-\pfail) (1-(1-\ber)^\linkack),\]
whereas the probability of a \emph{success} is
\[
\psucc\define (1-\pfail) (1-\ber)^{\linkack}.
\]

\noindent Note that we assume that acknowledgement frames are sent \emph{without} adding  redundancy. This is a reasonable assumption, given the current practices in most wireless technologies.

In order to estimate the energy consumption, we compute the expected number of bits sent at the link layer. Remark that in cases of success or partial failure, $\linkdata+\linkack$ bits are sent, whereas only $\linkdata$ bits are sent in cases of failure. Recall however that the redundancy ratio determines the number of useful bits per frame, and thus the number of frames to send.

We now take into account the link-layer sending attempts, following the ARQ technique implemented by the MAC protocol. We denote by $\linkrmax$ the maximum number of attempts, after which the sender of a data frame considers the receiver is unreachable.

The \emph{reception} failure probability of the data frame after $\linkrmax{}$ attempts is simply the probability $\per{}$ of having $r$ successive failures, i.e.:
\begin{equation}
\per{} \define \pfail^{\linkrmax{}}.
\label{eq:per_ber_rmax}
\end{equation}
In that case, the total number of bits sent only comes from the sender (no acknowledgement is sent), and thus equals 
\begin{equation}
\HS{}_{f}\define \linkrmax\times\linkdata.
\label{eq:linklayerfail}
\end{equation}

With probability $1-\per{}$, the receiver gets the data frame within the $\linkrmax$ link-layer attempts. In that case, the total number of bits sent depends on the number of failures and partial failures before a success, if any (we only know here that at least one attempt led to a success or a partial failure). There are two possibilities:
\begin{itemize}
\item either all $r$ attempts were failures or partial failures, but at least one was a partial failure (since we are in the case where the receiver got the data),
\item or the last of the $k \leq \linkrmax$ attempts was a success (in the sense of case (c) in Fig.~\ref{fig:link_layer_losses_success}). 
\end{itemize}

Conditioned on the receiver getting the data frame within the $\linkrmax$ link-layer attempts, the expected total number of bits sent $\HS{}_s$ can therefore be computed as follows:

\begin{equation}
\HS{}_{s} = \dfrac{1}{1 - \per{}}( 
\sum_{i=1}^{\linkrmax}\comb{i}{\linkrmax}\ppartial^i\pfail^{\linkrmax-i}(\linkrmax\linkdata+i\linkack) +\sum_{k=1}^{\linkrmax{}} \psucc \sum_{i=0}^{k-1}\comb{i}{k-1} \ppartial^i\pfail^{k-1-i}(k\linkdata+(i+1)\linkack)) .\label{eq:linklayersuccess}
\end{equation}

In~(\ref{eq:linklayersuccess}), $k$ stands for the index of the first success in $r$ attempts (if any), and $i$ is the number of partial failures among all attempts.

\subsection{Multi-hop model}
\label{sec:multi_hop_models}

Let us now focus on the multi-hop case. An end-to-end transmission succeeds if the message reaches the destination after a certain number $\hop$ of hops.

In this paper, we assume that link layer transmissions on each hop are independent. 
We denote $\ber{}_{i}$ the bit error rate and $\per{}_{i}$ the frame error rate of the $i^{th}$ hop.
$\per{}_{i}$ is computed as per (\ref{eq:per_ber_rmax}), taking $\ber = \ber{}_{i}$.
Therefore, the probability that a frame is correctly received by a destination node is simply the probability $\PEE{}_{s}$ that all $\hop$ one-hop transmissions succeed, i.e., 
\begin{equation}
\PEE{}_{s}= \prod_{i=1}^{\hop{}} (1-\per{}_{i})
\label{eq:p_ees}
\end{equation}
where $\hop{}$ is the number of hops from the sender to the receiver. 

We assume here that the MAC layer is able to detect duplicate frames, such as those that are produced in case of a partial failure (case (b) in Fig.~\ref{fig:link_layer_losses_success}). This can be implemented, for instance, by using a sequence number in the MAC frame headers; when a node receives from one neighbor two successive frames with the same sequence number, it assumes that the corresponding Ack frame was lost and deletes the second frame\footnote{Extending the current model to consider the case where there is no duplicate frame detection is left as future work. Note that the current 802.15.4 specification does not explicitly mandate such a procedure, though it is possible to implement it.}. This avoids the propagation of several copies of the same data frame.

Figure~\ref{fig:multi_hop_losses_success} shows the two possibilities for the outcome of the end-to-end transmission of a frame.

\gdef\lexscale{1.3}
\begin{figure}[ht]
\begin{center}
\subfigure[End-to-end failure scenario: the frame cannot be forwarded after $r$ unsuccessful retransmissions.]{
	\begin{tikzpicture}[xscale=\lexscale]
	\draw [very thick] (-2,-1) -- (-2,1) node [above] {Sender} ;
	\draw (-1,-1) -- (-1,1);
	\draw (0,-1) -- (0,1);
	\draw (1,-1) -- (1,1);
	\draw [very thick] (2,-1) -- (2,1) node [above] {Receiver} ;
	\draw [thick] [->] (-2,0.5) --  (-1,0.30) ;
	\draw [thick] [->] (-1,0.30) --  (-2,0.25) ;
	\draw [thick] [->] (-1,0.25) -- (0,0.05);
	\draw [thick] [->] (0,0.05) --  (-1,0.0) ;
	\draw [thick] [->] (0,0) --  (0.8,-0.2) ;
	\node [cross out,draw=red] at (0.8,-0.2) {.};
	\draw [thick] [->] (0,-0.35) --  (0.6,-0.50) ;
	\node [cross out,draw=red] at (0.6,-0.50) {.};
	\draw [thick] [->] (0,-0.65) --  (0.5,-0.78) ;
	\node [cross out,draw=red] at (0.5,-0.78) {.};
	\end{tikzpicture}
}
\subfigure[End-to-end success scenario: the frame arrives at the destination. This scenario may also include partial failures over one or more hops (not depicted).]{
	\begin{tikzpicture}[xscale=\lexscale]
	\draw [very thick](-2,-1) -- (-2,1) node [above] {Sender} ;
	\draw (-1,-1) -- (-1,1);
	\draw (0,-1) -- (0,1);
	\draw (1,-1) -- (1,1);
	\draw [very thick] (2,-1) -- (2,1) node [above] {Receiver} ;
	\draw [thick] [->] (-2,0.5) --  (-1,0.30) ;
	\draw [thick] [->] (-1,0.30) --  (-2,0.25) ;
	\draw [thick] [->] (-1,0.25) -- (0,0.05);
	\draw [thick] [->] (0,0.05) --  (-1,0.0) ;
	\draw [thick] [->] (0,0) --  (0.8,-0.2) ;
	\node [cross out,draw=red] at (0.8,-0.2) {.};
	\draw [thick] [->] (0,-0.3) --  (1,-0.50) ;
	\draw [thick] [->] (1,-0.5) --  (0,-0.55) ;
	\draw [thick] [->] (1,-0.55) --  (2,-0.75) ;
	\draw [thick] [->] (2,-0.75) --  (1,-0.80) ;
	\end{tikzpicture}
} 
\end{center}
\caption{Failure and success scenarios in a multi-hop transmission.}
\label{fig:multi_hop_losses_success}
\end{figure}
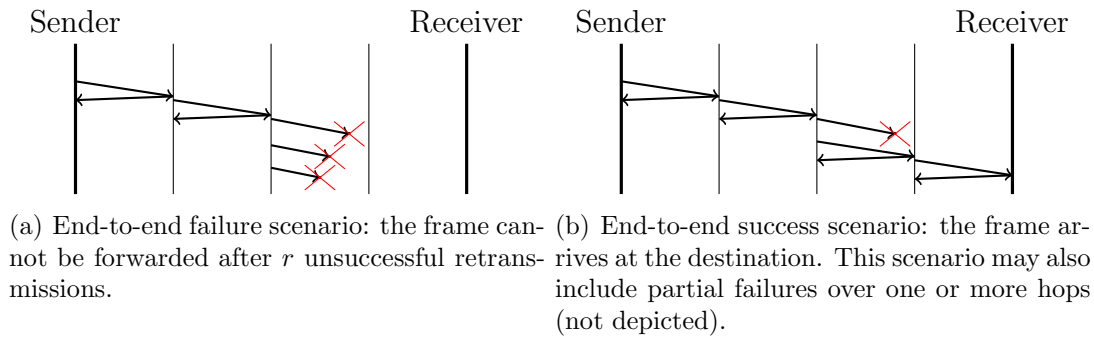

Again, we express the expected number of bits sent, conditioned on the success of the end-to-end transmission (encompassing possible link-layer retransmissions, within the limit of $\linkrmax$ total attempts per hop).
\begin{itemize}
\item Knowing that the destination node correctly receives the message, the expected number of bits sent by all network nodes is simply
\begin{equation}
\label{eq:ee_succ}
\EE{}_{s}\define \sum_{i=1}^{\hop{}}  \HS{}_{s_{i}}.
\end{equation}
\item Knowing that the message was lost in one of the $\hop$ hops, the  number of bits sent depends on the hop where the loss (i.e., the failure of all $\linkrmax$ attempts) occurs. The expected value then equals\vspace{-1ex}
\end{itemize}
\begin{eqnarray}
\EE{}_{f}\!\!\! &\define&\!\!\!\dfrac{ \sum_{k=1}^{\hop{}} (\sum_{i=1}^{k-1} \HS{}_{s_{i}}+ \HS{}_{f_{k}} )  \prod_{j=1}^{k-1} (1-\per{}_{j})  \per{}_{k} } {1 - \PEE{}_s}  
\label{eq:ee_fail}
\end{eqnarray}
\noindent where $\HS{}_{s_{i}}$ and $\HS{}_{f_{i}}$ are computed using (\ref{eq:linklayersuccess}) and (\ref{eq:linklayerfail}), respectively.

\subsection{TCP performance}
\label{sec:tcp_models}

We now study the energy consumption of TCP in a multi-hop wireless network, taking into account the fact that TCP segments may be fragmented by lower layers if the total size of data frames (that depends on the TCP maximum segment size (MSS)) exceeds the MTU of those layers.

We can intuitively think of several opposite effects of the MSS, which suggest that a trade-off has to be found:

\noindent\emph{i)} If TCP segments are fragmented, the loss of one fragment leads to the loss of the original TCP segment and therefore a new TCP end-to-end retransmission. Further, using short TCP segments (i.e., a small MSS) saves CPU power associated to fragmentation and reassembly. 
Moreover, sending several fragments (treated independently by lower layers) increases the probability of collision between two TCP data fragments, due to the hidden-node problem.
As an illustration of that phenomenon, one can refer to Fig.~\ref{fig:comp_collisions}, provided later on, where the number of collisions with long TCP segments is much larger than when short TCP segments are used.
\\\noindent\emph{ii)} On the other hand, the use of long TCP segments reduces the number of TCP segments sent by the source, and thus, the TCP header overhead (which can be very large, for small MTU values) and the number of TCP acknowledgements. 

To analyze such trade-off, we now apply the model presented in Section~\ref{sec:multi_hop_models}. Our objective here is to determine an MSS choice that optimizes TCP performance, in terms of energy consumption.

Let us consider that the MSS of TCP and the MTU of the MAC layer are such that each TCP segment is fragmented into $\nfrag$ link-layer frames. This encompasses the particular case when no fragmentation occurs, that simply corresponds to $\nfrag=1$. We will make the assumption that TCP ACKs always fit in a single MAC frame, though it would be easy to consider a more general case in which TCP ACKs are also fragmented. Note finally that we do not take into account the use of the delayed ACKs mechanism by the TCP receiver; including the effect of such a mechanism is left to future work. 

To simplify the analysis, we assume that all fragments of a TCP data segment have the same total size $\linkdata_{\text{DATA}}$, and  all TCP acknowledgement frames have the same total size $\linkdata_{\text{ACK}}$.
At the destination, the segment is reconstructed from the received fragments. The loss of a single link-layer frame induces the loss of the whole TCP segment, and thus the retransmission of all $\nfrag$ fragments. Further, we consider that the number of TCP retransmissions is not limited; that is, the TCP source keeps on sending a segment until it receives a TCP acknowledgement from the TCP receiver.

From the previous analysis, the success probability $P_s$ of  a TCP segment transmission attempt is simply the probability that all $\nfrag$ data fragments be correctly sent to the destination, and the TCP ACK be successfully sent back to the source:
\begin{equation*}
P_s = \PEE{}_{s}^\nfrag \times \PEE{}_{s,\text{ack}} \; ,
\end{equation*}
\noindent where $\PEE{}_{s}$ and $\PEE{}_{s,\text{ack}}$ denote the success probability of the multi-hop transmission of a TCP data fragment and of a TCP acknowledgement frame, respectively.
$\PEE{}_{s}$ and $\PEE{}_{s,\text{ack}}$ are simply obtained by applying~(\ref{eq:p_ees}),
but replacing $\linkdata$ by respectively $\linkdata_{\text{DATA}}$ and $\linkdata_{\text{ACK}}$.

In our model, each TCP data fragment transmission succeeds or fails independently of the others. Knowing that a transmission is successful \emph{at the TCP level} (i.e., the TCP ACK is correctly received by the TCP source, which implies that all $\nfrag$ fragments correctly reached the destination), the expected total number of bits sent by all nodes equals:
\begin{equation*}
\Seg{}_{s}\define \EE{}_{s} \times \nfrag + \EE{}_{s,\text{ack}}
\end{equation*}
As above,  $\EE{}_{s}$ and $\EE{}_{s,\text{ack}}$ are obtained from~(\ref{eq:ee_succ}) using $\linkdata_{\text{DATA}}$ and $\linkdata_{\text{ACK}}$, respectively, as the size of a frame.
 Likewise, we also define $\EE{}_{f,\text{ack}}$ from~(\ref{eq:ee_fail}).

Knowing that a TCP transmission attempt has failed, the expected number of bits sent end-to-end  by all nodes is:
\begin{equation*}
\Seg{}_{f} \define \dfrac{1}{1-P_{s}}\Big[\underbrace{\TEE{}_{f}(1- \PEE{}_{s}^\nfrag )}_{\substack{\text{end-to-end transmission failure} \\ \text{of one or more of the $\nfrag$ fragments}}}+\underbrace{(\EE{}_{s} \times \nfrag +\EE{}_{f,\text{ack}})  \PEE{}_{s}^\nfrag  (1- \PEE{}_{s,\text{ack}} )}_{\text{end-to-end transmission failure of the TCP ACK}}\ \Big],
\end{equation*}
where $\TEE{}_{f}$ is the expected total number of bits sent for the $\nfrag$ fragments to reach the destination
, knowing that they (i.e., at least one) finally fail.
Formally,
\[
\TEE{}_{f}
\define 
\sum_{k=1}^{\nfrag}  {\nfrag \choose k} \big( k \EE{}_{f} + (\nfrag-k)
 \EE{}_{s}\big) (1-\PEE{}_{s})^k (\PEE{}_{s})^{\nfrag-k}.  
\]
After some algebra, we get
\[
\TEE{}_{f}= \nfrag (1-\PEE{}_{s}) \EE{}_{f}+ \nfrag  \EE{}_{s} \PEE{}_{s}  (1- \PEE{}_{s}^\nfrag).
\]

To simplify the analysis, we will assume that the TCP window is equal to one TCP segment. This assumption is justified because a small window is a sensible choice for networks with a moderate number of hops \cite{FuLZLZG05}. Moreover, such small windows are typical of current TCP implementations for LLNs (e.g., Contiki OS \cite{contiki}), since the memory and CPU constraints of wireless embedded devices make it difficult to fully implement TCP's congestion control mechanisms.

We can now compute the total number of bits that have to be sent by all nodes, to successfully transmit both a TCP segment and its corresponding TCP ACK. Since we assumed the number of TCP retransmissions is unbounded (i.e., the TCP sender keeps on retrying until the transmission succeeds), the mean number of transmissions for a given TCP segment equals $1/P_s$. 
 This therefore corresponds to a total number of bits sent (per segment) of
\[
\Seg{} \define  \Seg{}_{f} (1/P_{s} -1) + \Seg{}_{s}.
\]


Finally, to successfully send a given amount $M$ of \emph{application} (``useful'') data, the expected number of bits sent by all wireless nodes is the number of segments $\lceil M/\text{MSS}\rceil$, multiplied by the expected number of bits sent for a successful segment transmission.
It is that final value that will be considered as representative of the overall energy consumption of the TCP transmission.

\section{Results and discussion}
\label{sec:analytical_simulation_result}

We will now compare the predictions of our analytical model to simulation results, and discuss the tradeoff between sending long or short segments in different scenarios.

We apply our model to study the energy consumption of TCP on IPv6 over Low-Power Wireless Personal Area Networks~\cite{rfc:lowpan}, considering IEEE 802.15.4~\cite{ieee802154} as the link-layer technology. The maximum link-layer frame size is thus equal to 127~bytes.
The 6LoWPAN layer adapts the IPv6 datagrams to the link-layer MTU, as explained in~\cite{rfc:lowpan}.

We consider two MSS choices (MSS=$64$ and MSS=$512$ bytes) for a given TCP session.
When the MSS is $64$ bytes, no fragmentation is performed by the 6LoWPAN layer, whereas in the case where MSS=$512$ bytes, this adaptation layer splits each segment into $8$ frames. 

\subsection{Energy model}

\newcommand{\neighbors}{n}
To compute the energy consumed by wireless nodes to successfully send a data frame, we specify in this section a simple energy model. We denote by $\neighbors$ the average number of neighbors of each node\footnote{In the plotted curves, we considered $\neighbors=2$.}. If a frame is sent by a wireless node, it is received by its $\neighbors$ neighbors (assumed thus to be in an active listening state). Therefore the energy consumed to send a link-layer frame equals
\[
E \define (\text{TransmitEnergy}+\neighbors \times \text{ReceiveEnergy}) \times \text{FrameSize} 
\]
Table~\ref{tab:energy_parameters} shows the energy consumption values that we considered in our numerical computations.
\begin{table}[ht]
\centering
\caption{Energy parameters.\label{tab:energy_parameters}}
\begin{tabular}{|l||c|}
\hline \textbf{Parameter} &  \textbf{Value}\\ 
\hline Transmit Energy  &  0.24 $\mu$J/bit\\ 
\hline Receive Energy &  0.21 $\mu$J/bit\\
\hline 
\end{tabular} 
\end{table}
Those values correspond to a transmit and receive consumption currents 
of $20$ mA and $17.7$ mA, respectively, a tension of $3$V and a transmission rate of $250$ kbit/s (thus, a per-bit transmission duration of $4$~$\mu$s). Such figures are typical of low-power wireless modules currently on the market (e.g., Crossbow TelosB\footnote{http://www.hoskin.qc.ca/uploadpdf/Instrumentation/CrossBow/hoskin\_TPR2400CA\_42efb73715b8b.pdf})
.

\subsection{Model assessment}

We first validate the results of our analytical model, through the simulation of the following scenario. A TCP sender sends a short file (51.2 Kbytes) to a TCP receiver.
We consider an average size of 20 bytes for IPv6 headers, thanks to the $\text{LOWPAN\_IPHC}$~\cite{rfc:lowpan_iphc} compression.
The simulation results plotted are average values after 30~simulation runs. The TCP window size is set to $1$, which implies that no congestion losses occur (we only consider one flow here). 
As a result, retransmissions can only be due to transmission errors, as described in Section~\ref{sec:energy_models}. For the sake of simplicity, all links have the same bit-error rate $B_{i} = B, i = 1,\ldots,h$. Unless indicated otherwise, the parameters used in simulations and in numerical computations correspond to the default values shown in Table~\ref{tab:sim_par}. Finally, as in Section~\ref{sec:tcp_models}, the maximum number of transport-layer retransmissions is not bounded.

\begin{table}[ht]
\centering
\caption{Default simulation parameters\label{tab:sim_par}}
\begin{tabular}{|l||c|}
\hline \textbf{Parameter} &  \textbf{Value}\\ 
\hline $\hop$ &  5\\ 
\hline $\linkrmax$ & 3\\
\hline $\frr$ & 0\\
\hline BER $\ber$& $3 \times 10^{-4}$\\
\hline Link-layer Ack frame size & 40 bits\\
\hline Link-layer data frame header & 120 bits\\
\hline IP header & 160 bits\\
\hline TCP header & 160 bits\\
\hline 
\end{tabular} 
\end{table}

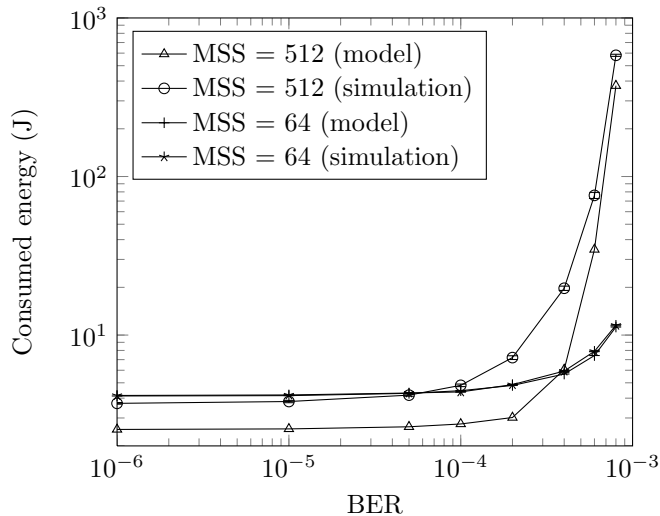
\begin{figure}[ht]
\begin{center}
{\footnotesize
\begin{tikzpicture}
\begin{loglogaxis}[legend style={at={(0.03,0.97)},anchor=north west},
xlabel=BER,
ylabel=Consumed energy (J),
ymin=2,ymax=1e3,
xmin=1e-6,xmax=1e-3,
every axis legend/.append style={nodes={right}}, 
]

\addplot  [mark= triangle]  plot 
coordinates {
(1e-6,2.54) 
(1e-5,2.56) 
(5e-5,2.64) 
(1e-4,2.75)
(2e-4,3.02)
(4e-4,6.06)
(6e-4,34.67)
(8e-4,374) 
};
\addlegendentry{MSS = 512 (model)}

\addplot  [mark=o] plot[error bars/.cd,
y dir=both,y explicit]
coordinates {
(1e-6,3.71) +- (0,0.044)
(1e-5,3.81) +- (0,0.065)
(5e-5,4.18) +- (0,0.064)
(1e-4,4.82) +- (0,0.089)
(2e-4,7.21) +- (0,0.196)
(4e-4,19.74) +- (0,0.618)
(6e-4,76.05) +- (0,3.37)
(8e-4,581.42) +- (0,8.93)
};
\addlegendentry{MSS = 512 (simulation)}

\addplot [mark=+]  plot
coordinates {
(1e-6,4.13) 
(1e-5,4.15) 
(5e-5,4.29) 
(1e-4,4.46) 
(2e-4,4.81)  
(4e-4,5.68)  
(6e-4,7.43)  
(8e-4,11.24)
};
\addlegendentry{MSS = 64 (model)}

\addplot plot[error bars/.cd,
y dir=both,y explicit]
coordinates {
(1e-6,4.16) +- (0,0.01)
(1e-5,4.192) +- (0,0.01)
(5e-5,4.31) +- (0,0.03)
(1e-4,4.38) +- (0,0.0158)
(2e-4,4.87) +- (0,0.013)
(4e-4,5.93) +- (0,0.028)
(6e-4,7.90) +- (0,0.041)
(8e-4,11.53) +- (0,0.089)
};
\addlegendentry{MSS = 64 (simulation)}

\end{loglogaxis}
\end{tikzpicture}
}
\end{center}
\caption{Energy consumption with long or short TCP segments, as a function of the BER $B$.\label{fig:con_energy_short_long_mss_ber}}
\end{figure}

\begin{figure}[ht]
\begin{center}
{\footnotesize

\begin{tikzpicture}
\begin{semilogyaxis}[
x tick label style={
/pgf/number format/1000 sep=},
ylabel=Number of collisions,
xlabel=Node number,
enlargelimits=0.15,
legend style={at={(0.03,0.97)},
anchor=north west,legend columns=1},
ybar,
bar width=5pt,
ymax=1e4,
every axis legend/.append style={nodes={right}}, 
]
\addplot [color=gray,fill=black] plot
coordinates {
(1,2)
(2,14)
(3,56) 
(4,56) 
(5,7)
(6,2)
};
\addplot [draw=black] plot
coordinates {
(1,85)
(2,1572)
(3,1232)
(4,1034)
(5,106)
(6,58)
};
\legend{MSS = 64,MSS = 512}
\end{semilogyaxis}
\end{tikzpicture}
}
\end{center}
\caption{Number of collisions in a multi-hop scenario.\label{fig:comp_collisions}}
\end{figure}
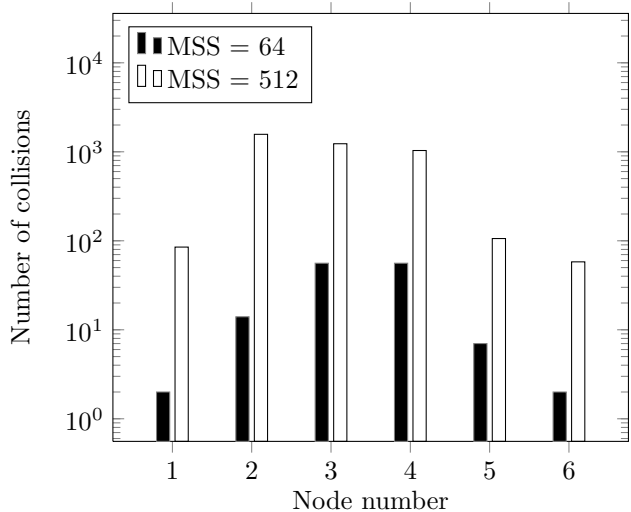

Figure~\ref{fig:con_energy_short_long_mss_ber} shows that analytical results closely match simulations results when a short TCP segment size is chosen.
However, we can see that 
with long TCP segments, 
the model tends to underestimate the energy consumption.
This difference comes from the presence of collisions between fragments of a given TCP data segment, 
that are not taken into account in the analytical model but do occur in simulation, as illustrated in Fig.~\ref{fig:comp_collisions}. In the figure, nodes 1 and 6 are the source and destination TCP nodes, respectively, and node 2 to 5 are intermediate nodes.

Moreover, Fig.~\ref{fig:con_energy_short_long_mss_ber} 
shows that using short TCP segments becomes interesting, from an energy point of view, when the bit-error rate is high (above $10^{-4}$).
For example, for a BER of $4\times10^{-4}$, the total consumed energy with MSS $= 512$ bytes is three times larger than with MSS $= 64$ bytes. For low values of $\ber$, the number of retransmissions is small, hence the energy consumption becomes roughly independent of $\ber$.

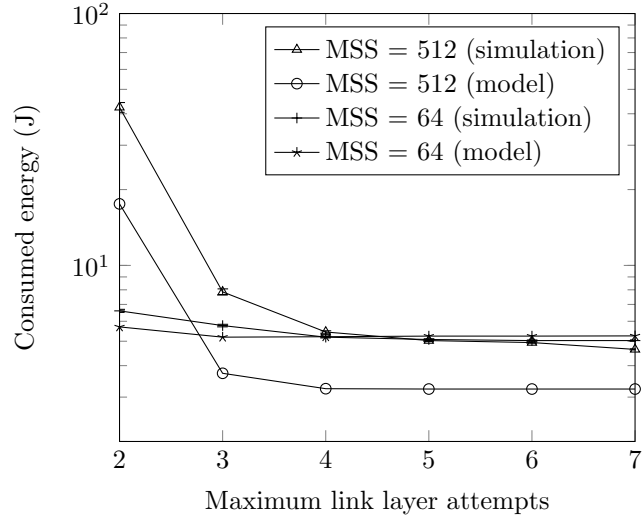
\begin{figure}
\begin{center}
{\footnotesize
\begin{tikzpicture}
\begin{semilogyaxis}[legend style={at={(0.97,0.97)}, anchor=north east},
xlabel=Maximum link layer attempts,
ylabel=Consumed energy (J),
ymin=2,ymax=100,
xmin=2,xmax=7,
every axis legend/.append style={nodes={right}}, 
]
 
\addplot [mark=triangle]  plot[error bars/.cd,
y dir=both,y explicit]
coordinates {
(2,42.43) +- (0,1.89)
(3,7.85) +- (0,0.2193)
(4,5.44) +- (0,0.061)
(5,5.03) +- (0,0.045)
(6,4.94) +- (0,0.031)
(7,4.64) +- (0,0.023)};
\addlegendentry{MSS = 512 (simulation)}

\addplot [mark=o] plot
coordinates {
(2,17.55) 
(3,3.73) 
(4,3.24) 
(5,3.23) 
(6,3.23)
(7,3.23) };
\addlegendentry{MSS = 512 (model)}

\addplot  [mark=+] plot[error bars/.cd,
y dir=both,y explicit]
coordinates {
(2,6.60) +- (0,0.093)
(3,5.77) +- (0,0.067)
(4,5.19) +- (0,0.037)
(5,5.08) +- (0,0.035)
(6,5.03) +- (0,0.027)
(7,5.03) +- (0,0.029)};
\addlegendentry{MSS = 64 (simulation)}

\addplot  plot
coordinates {
(2,5.70) 
(3,5.19) 
(4,5.21) 
(5,5.24) 
(6,5.24) 
(7,5.25)};
\addlegendentry{MSS = 64 (model)}

\end{semilogyaxis}
\end{tikzpicture}
}
\end{center}
\caption{Energy consumption with short or long TCP segments, as a function of the number of link layer attempts $\linkrmax$ (with $B=5 \times 10^{-4}$).\label{fig:con_energy_short_long_mss_rmax}}
\end{figure}

We remark on Figs.~\ref{fig:con_energy_short_long_mss_ber} and~\ref{fig:con_energy_short_long_mss_rmax} that simulation results fit better our analytical results when no fragmentation is performed by the link layer. This is again due to collisions that occur in the large-MSS case and are not encompassed by our model.
Fig.~\ref{fig:con_energy_short_long_mss_rmax} also illustrates that 
increasing the number of link-layer attempts decreases the 
energy consumption, especially so when there is fragmentation (i.e., a large MSS).
Indeed, giving the link layer more chances to send a data frame to its next-hop node  reduces the discard probability of TCP segments, and therefore the number of end-to-end retransmissions.
%

In the following, we study the impact of the other system parameters on the energy consumption, based on the model only. We should therefore stay aware that the energy consumption may be a little underestimated for large values of the MSS.

\subsection{FEC redundancy ratio and energy consumption}

%
%
%

\begin{figure}
\begin{center}
{\footnotesize
\begin{tikzpicture}
\begin{loglogaxis}[legend style={at={(0.97,0.97)}, anchor=north east},
xlabel=Redundancy ratio ($\frr$),
ylabel=Consumed energy (J),
ymin=1,ymax=1000,
xmin=0.001,xmax=1,
every axis legend/.append style={nodes={right}}, 
]
 
\addplot  [mark=none,thick,black] 
coordinates {
(0.001,50110) 
(0.002,50738) 
(0.003,10.1) 
(0.004,10.3) 
(0.005,2.67) 
(0.006,2.67) 
(0.007,2.67) 
(0.008,2.66) 
(0.009,2.66) 
(0.01,2.57) 
(0.02,2.59) 
(0.03,2.61) 
(0.04,2.63) 
(0.05,2.57)
(0.06,2.68)
(0.07,2.70)
(0.08,2.72)
(0.09,2.75)
(0.1,2.77)
(0.2,3.00)
(0.3,3.23)
(0.4,3.46)
(0.5,3.69)
(0.6,3.92)
(0.7,4.15)
(0.8,4.38)
(0.9,4.61)
(1,4.85)};
\addlegendentry{MSS = 512, $\linkrmax$=1}

\addplot  [mark= none,thick,black,densely dashed] 
coordinates {
(0.001,3.02) 
(0.002,3.21) 
(0.003,2.48) 
(0.004,2.49) 
(0.005,2.43) 
(0.006,2.44) 
(0.007,2.44) 
(0.008,2.44) 
(0.009,2.44) 
(0.01,2.44) 
(0.02,2.47) 
(0.03,2.49) 
(0.04,2.49) 
(0.05,2.51)
(0.06,2.56)
(0.07,2.58)
(0.08,2.61)
(0.09,2.63)
(0.1,2.65)
(0.2,2.89)
(0.3,3.12)
(0.4,3.35)
(0.5,3.58)
(0.6,3.82)
(0.7,4.05)
(0.8,4.28)
(0.9,4.52)
(1,4.75)};
\addlegendentry{MSS = 512, $\linkrmax$=3}

\addplot [mark=none,thick,black, densely dotted]
coordinates {
(0.001,16.51) 
(0.002,16.54) 
(0.003,7.61) 
(0.004,7.62) 
(0.005,4.29) 
(0.006,4.29) 
(0.007,4.29) 
(0.008,4.30) 
(0.009,4.30) 
(0.01,4.17) 
(0.02,4.20) 
(0.03,4.23) 
(0.04,4.27) 
(0.05,4.31)
(0.06,4.34) 
(0.07,4.38) 
(0.08,4.42) 
(0.09,4.46) 
(0.1,4.49)
(0.2,4.86)
(0.3,5.23)
(0.4,5.60)
(0.5,5.97)
(0.6,6.34)
(0.7,6.71)
(0.8,7.08)
(0.9,7.45)
(1,7.82)};
\addlegendentry{MSS = 64, $\linkrmax$=1}

\addplot  [mark=none,thick,black, loosely dashed] 
coordinates {
(0.001,5.20) 
(0.002,5.19) 
(0.003,4.53) 
(0.004,4.53) 
(0.005,4.26) 
(0.006,4.26) 
(0.007,4.26) 
(0.008,4.26) 
(0.009,4.27) 
(0.01,4.26) 
(0.02,4.30) 
(0.03,4.34) 
(0.04,4.37) 
(0.05,4.41)
(0.06,4.45) 
(0.07,4.49) 
(0.08,4.53) 
(0.09,4.56) 
(0.1,4.60)
(0.2,4.98)
(0.3,5.36)
(0.4,5.74)
(0.5,6.12)
(0.6,6.50)
(0.7,6.88)
(0.8,7.26)
(0.9,7.63)
(1,8.01)};
\addlegendentry{MSS = 64, $\linkrmax$=3}

\end{loglogaxis}
\end{tikzpicture}
}
\end{center}
\caption{Consumed energy using short or long TCP segment, as a function of the redundancy ratio $\alpha$ ($\ber=3 \times 10^{-4},\hop=5$).\label{fig:con_energy_redundancy_rmax}}
\end{figure}
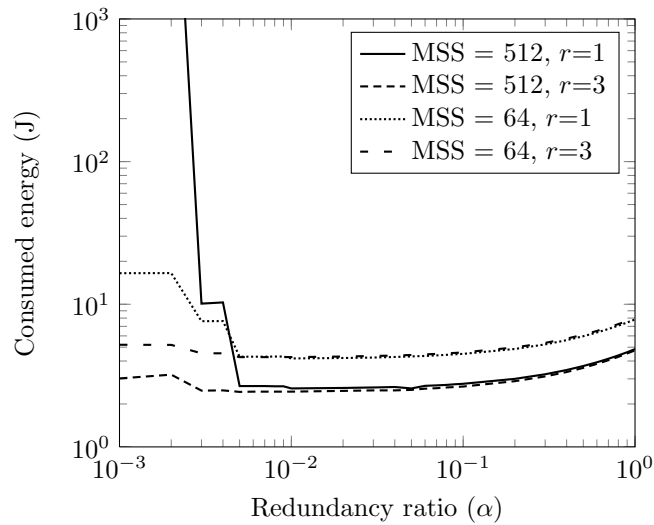

We first study the impact of FEC on the energy consumption. Fig.~\ref{fig:con_energy_redundancy_rmax} 
shows that there seems to be an optimal amount of redundancy, in terms of energy consumption. Below the optimal value, adding redundancy reduces the probability of losses, and thus reduces the energy consumption. When $\alpha$ is above such optimal value, however, energy expenditure steadily increases because of the redundancy overhead.

Recall that the MTU being fixed, the number of frames to send depends on the redundancy ratio according to a stairstep function, hence the discontinuities in the figure.

\subsection{Selecting the TCP MSS to minimize energy consumption}
\label{sec:select_tcp_mss_rmax}

Depending of the numerous parameters of a given scenario, it appears that the same MSS value is not always the most efficient one in terms of energy. We now intend to summarize the effect of all parameters, by focusing on the best MSS strategy to implement. 
In the following figures, we compare the two values $MSS=64$ and $MSS=512$, and we concentrate on the boundary values, that delimitate zones where one MSS value outperforms the other.

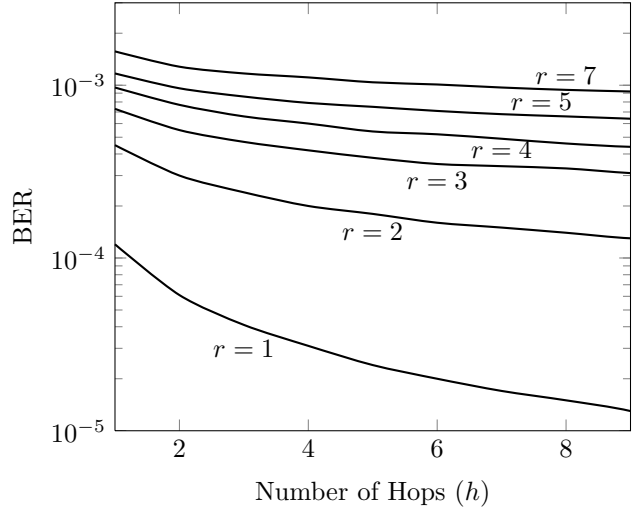
\begin{figure}[ht]
\begin{center}
{\footnotesize
\begin{tikzpicture}
\begin{semilogyaxis}[xlabel=Number of Hops ($\hop$), ylabel=BER,
xmin=1, xmax=9, ymin=1e-5,ymax=3e-3,enlargelimits=false,
every axis legend/.append style={nodes={right}}, 
]
\draw (axis cs:3,3e-5)node{$\linkrmax=1$};
\draw (axis cs:5,1.8e-4)node[anchor=north]{$\linkrmax=2$};
\draw (axis cs:6,2.8e-4)node{$\linkrmax=3$};
\draw (axis cs:7,5.4e-4)node[anchor=north]{$\linkrmax=4$};
\draw (axis cs:7,8 e-4)node[anchor=west]{$\linkrmax=5$};
\draw (axis cs:8,9e-4)node[anchor=south]{$\linkrmax=7$};
\addplot[mark=none,thick,smooth, black]
coordinates {
(1,0.00012) 
(2,0.000061) 
(3,0.000041) 
(4,0.000031) 
(5,0.000024) 
(6,0.000020) 
(7,0.000017) 
(8,0.000015) 
(9,0.000013) 
(10,0.00001) };
\addplot[mark=none,thick,smooth, black]
coordinates {
(1,0.00045) 
(2,0.00030) 
(3,0.00024) 
(4,0.00020) 
(5,0.00018) 
(6,0.00016) 
(7,0.00015) 
(8,0.00014) 
(9,0.00013) 
(10,0.00013) };
\addplot[mark=none,thick,smooth, black]
coordinates {
(1,0.00073) 
(2,0.00055) 
(3,0.00047) 
(4,0.00042)
(5,0.00038) 
(6,0.00035) 
(7,0.00034) 
(8,0.00033) 
(9,0.00031) 
(10,0.0003)  };
\addplot[mark=none,thick,smooth, black]
coordinates {
(1,0.00097) 
(2,0.00077) 
(3,0.00066)
(4,0.00060) 
(5,0.00054)
(6,0.00052) 
(7,0.00049)
(8,0.00046) 
(9,0.00044)
(10,0.00044) };
\addplot[mark=none,thick,smooth, black]
coordinates{
(1,0.00117) 
(2,0.00096) 
(3,0.00086) 
(4,0.00079) 
(5,0.00075) 
(6,0.00071) 
(7,0.00068)
(8,0.00066) 
(9,0.00064)
(10,0.00062)};
\addplot[mark=none,thick,smooth, black]
coordinates {
(1,0.00157) 
(2,0.00128) 
(3,0.00117) 
(4,0.00111) 
(5,0.00104) 
(6,0.00101) 
(7,0.00097) 
(8,0.00094) 
(9,0.00092)
(10,0.0009)};
\end{semilogyaxis}
\end{tikzpicture}
}
\end{center}
\caption{Long (MSS=$512$ bytes) versus short (MSS=$64$ bytes) in a multi-hop TCP transmission: prefer the short MSS above the curves, the long one below. \label{fig:tradeoff_log_short_max_attempts}}
\end{figure}
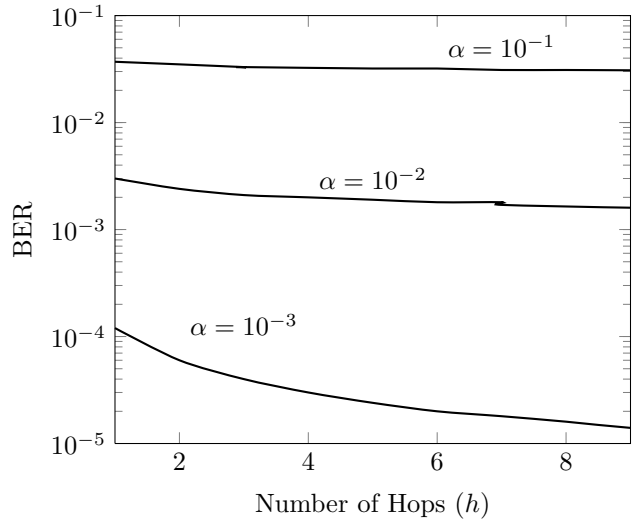
\begin{figure}[ht]
\begin{center}
{\footnotesize
\begin{tikzpicture}
\begin{semilogyaxis}[xlabel=Number of Hops ($\hop$), ylabel=BER,
xmin=1, xmax=9, ymin=1e-5,ymax=1e-1,enlargelimits=false,
every axis legend/.append style={nodes={right}}, 
]
\draw (axis cs:3,2e-4)node[anchor=north]{$\frr=10^{-3}$};
\draw (axis cs:5,3e-3)node{$\frr =10^{-2}$};
\draw (axis cs:7,8e-2)node[anchor=north]{$\frr =10^{-1}$};
\addplot[mark=none,thick,smooth, black]
coordinates {
(1,0.00012) 
(2,0.00006) 
(3,0.00004) 
(4,0.00003) 
(5,0.000024) 
(6,0.00002) 
(7,0.000018) 
(8,0.000016) 
(9,0.000014)
(10,0.000013) };
\addplot[mark=none,thick,smooth, black]
coordinates {
(1,0.0030) 
(2,0.0024) 
(3,0.0021) 
(4,0.0020) 
(5,0.0019) 
(6,0.0018) 
(7,0.0018) 
(7,0.0017) 
(9,0.0016)
(10,0.0016) };
\addplot[mark=none,thick,smooth, black]
coordinates {
(1,0.037) 
(2,0.035) 
(3,0.033)
(3,0.033) 
(5,0.032) 
(6,0.032) 
(7,0.031) 
(8,0.031) 
(9,0.0308)
(10,0.0306)};
\end{semilogyaxis}
\end{tikzpicture}
}
\end{center}
\caption{Long (MSS=$512$ bytes) versus short (MSS=$64$ bytes)
in a multi-hop TCP transmission: prefer the short MSS above the curves, the long one below. \label{fig:tradeoff_log_short_fec}}
\end{figure}

We  plot those frontier curves in Figs.~
\ref{fig:tradeoff_log_short_max_attempts} and~\ref{fig:tradeoff_log_short_fec}. In each figure, the area above the curve represents the case where a TCP MSS value of $64$ bytes consumes less energy than  an MSS of $512$ bytes, while the opposite holds below the curves.



Figure~\ref{fig:tradeoff_log_short_max_attempts} shows the two zones, depending on the transmission distance and the BER. We remark that for a given BER, short MSSs tend to outperform long MSSs when the distance grows: it is more and more interesting to use short MSS values instead of long ones. Indeed, for large networks the cost of end-to-end retransmissions will exceed the potential economy in terms of TCP header overhead. For a given distance, short MSSs are better suited for large BER environments since (energy-spending) segment retransmissions  tend to occur more frequently. As previously observed, increasing the maximum number of link layer attempts $\linkrmax$ reduces the one-hop transmission failures, and thus limit the effects of errors, hence favoring long MSSs.

Finally, Fig.~\ref{fig:tradeoff_log_short_fec} illustrates the effect of FEC mechanisms. Not surprisingly (the FEC reducing the effect of transmission errors), redundancy makes large MSSs outperforms small MSSs due to the overhead reduction they allow.

\section{Conclusions and Future Work}
\label{sec:conclusion}

In this paper, we have proposed an analytical model to estimate the
number of bits sent by all wireless nodes in a TCP session in a Low power and Lossy Network, in order to evaluate the overall energy consumption.
The model has been validated through simulations, using the INETMANET framework\cite{inet} of the
OMNet++ network simulator, in the context of TCP over 6LoWPANs.

Our main outcomes regard the choice of an energy-saving Maximum Segment Size  (MSS) for TCP.
We have shown that using a large TCP segment size is less energy-consuming in small, low-error networks, while it becomes interesting to reduce the MSS when the network is large or very lossy. The impact of the number of attempts at the link layer, as well as the use of FEC, has also been studied.

The work presented here could be extended in several ways. A first interesting direction would be to model the collision process, that has been observed in our simulations for large MSS values. We would also like to consider the case when duplicate frames are not detected at the link layer; likewise, the transport layer modeling could be extended to encompass TCP's delayed acknowledgement mechanism, and also larger TCP windows. Finally, we are currently investigating 
an adaptation algorithm which dynamically adjusts the TCP segment size to the optimal MSS value.

\section{Acknowledgements}
\label{sec:acknowledgement}
The work of A.\ Ayadi has been funded by the P\^ole de Recherche Avanc\'ee en Communications (PRACom). We would like to thank Ahmed Triki for his help with the simulations.

\bibliographystyle{plain}
\bibliography{../biblioAA}

%
\end{document}